\documentclass[12pt]{article}
\usepackage{graphicx}
 \oddsidemargin  = - 7 mm 
 \evensidemargin = - 7 mm
\input{epsf.tex}

\def\jtstrut{\vrule height4ex depth0pt width0pt} 

\begin{document}
\date{}
\title{$D$ mesic nuclei} 
\author{
 {\large C. Garc\'{\i}a--Recio\thanks{
e-mail: g\_recio@ugr.es}} \\
 {\small Departamento de F\'{\i}sica At\'omica Molecular y Nuclear, 
  Universidad de Granada, 18071 Granada, Spain}\\
{\large J. Nieves}\\
{\small Departamento de F\'{\i}sica Te\'orica and IFIC, 
  Centro Mixto Universidad de Valencia-CSIC }\\
 {\small Institutos de Investigaci\'on de Paterna,
  Apdo. correos 22085, 46071, Valencia, Spain }\\
 {\large L. Tolos} \\
 {\small Theory Group, KVI, University of Groningen, Zernikelaan 25, 9747AA Groningen, The Netherlands }
       } 
\maketitle
\begin{abstract} 
 The energies and widths of several $D^0$ meson bound states for
 different nuclei are obtained using a $D$-meson selfenergy in the
 nuclear medium, which is evaluated in a selfconsistent manner using
 techniques of unitarized coupled-channel theory. The kernel of the
 meson-baryon interaction is based on a model that treats heavy
 pseudoscalar and heavy vector mesons on equal footing, as required by
 heavy quark symmetry. We find $D^0$ bound states in all studied
 nuclei, from $^{12}\mbox{C}$ up to $^{208}\mbox{Pb}$. The inclusion
 of vector mesons is the keystone for obtaining an attractive
 $D$-nucleus interaction that leads to the existence of $D^0$-nucleus
 bound states, as compared to previous studies based on SU(4) flavor
 symmetry.  In some cases, the half widths are smaller than the
 separation of the levels, what makes possible their experimental
 observation by means of a nuclear reaction. This can be of particular
 interest for the future ${\rm {\bar P}}$ANDA@FAIR physics program. We
 also find a $D^+$ bound state in $^{12}$C, but it is too broad and
 will have a significant overlap with the energies of the continuum.

\end{abstract}

\section{Introduction}

 The $D$-meson nucleus optical potential has been a subject of intense
 study over the last years. In particular, modifications of the $D$
 meson properties in an hadronic environment might influence the
 rhythm of charmonium production \cite{10,11,12,13,14,15,16}, as a
 complementary explanation for charmonium suppression in a Quark-Gluon
 Plasma \cite{matsui}. Moreover, $D$-meson bound states in $^{208}$Pb
 were predicted \cite{tsushima} relying upon a strong mass shift for
 $D$ mesons in the nuclear medium based on a quark-meson coupling
 model~\cite{sibirtsev}.  The experimental observation of these bound
 states might be however problematic since, even if they exist, their
 widths could be very large~\cite{tolos-angels-mizutani} as compared
 to the separation of the levels.  In
 Ref.~\cite{tolos-angels-mizutani}, the $D$-meson potential was
 obtained using techniques of self-consistent unitarized
 coupled-channel theory adapted to the meson-baryon
 interaction~\cite{Oset:1997it,oller}, which followed and extended the
 works using chiral Lagrangians and the Lippmann-Schwinger equation
 initiated in Refs.~\cite{wolfram,kaiser}. This work followed a
  scheme similar to those used in previous approaches on the spectral
 features of $D$ mesons in symmetric nuclear
 matter~\cite{tolos-schaffne-mishra,tolos-schaffne-stoecker,korpa-lutz,tetsuro-angels}. The
 systematic inclusion of medium corrections to the scattering equation
 is crucial for the generation and modification of the
 $\Lambda_c(2595)$ resonance in the nuclear medium and, thus,
 eliminate the main source of uncertainty of earlier evaluations of
 the $D$ nucleus optical potential.  In fact, this has been also done
 for the $\eta$-meson selfenergy in a nuclear medium in
 Ref.~\cite{Inoue:2002xw}, using the vacuum amplitude of
 Ref.~\cite{inoue}. The $\eta$ nucleus optical potential was evaluated
 in a selfconsistent way, as it was done for the antikaon case in
 Ref.~\cite{ro}.  Later this optical potential was used for studying
 the possible $\eta$-nucleus bound states in
 Ref.~\cite{GarciaRecio:2002cu}, where the energy dependence of the
 $\eta$ selfenergy around the $\eta N$ threshold was taken into account
 and shown to be very much relevant for the problem.

Here we will undertake a similar study for the $D^0-$nucleus bound
states using a recently generated $D$ meson selfenergy in the nuclear
medium \cite{Tolos:2009nn}. This model incorporates heavy-quark
symmetry in the charm sector improving in this respect with the
recent $t$-channel vector meson-exchange approaches
\cite{tolos-angels-mizutani,korpa-lutz,tetsuro-angels,JimenezTejero:2009vq}.
As a consequence, the pseudoscalar $D$ meson and the $D^*$ meson, its
vector partner, are treated on equal footing. This new scheme
generates a broad spectrum of new resonant meson-baryon states in the
charm one and strangeness zero~\cite{GarciaRecio:2008dp} and the
exotic charm minus one~\cite{Gamermann:2010zz} sectors. Furthermore,
this framework allows to obtain simultaneously the properties of $D$
and $D^*$ mesons in nuclear matter~\cite{Tolos:2009nn}.  The in-medium
calculation of Ref.~\cite{Tolos:2009nn} includes Pauli blocking
effects on the nucleon and the $D$ and $D^*$ selfenergies in a
self-consistent manner. Moreover in this work, a novel renormalization
scheme is introduced that guaranties that the nuclear medium
corrections do not depend on the ultraviolet cutoff used to
renormalize the free space amplitudes. Compared to previous results
\cite{tolos-angels-mizutani,tetsuro-angels}, the width of the $D$
meson in nuclear matter is small with respect to the mass shift and,
therefore, bound states for $D$ mesons in nuclei might be expected.

This paper is organized as follows. In Sect.~\ref{sec2} we introduce the
$D$-nucleus potential from Ref.~\cite{Tolos:2009nn}. The results for
the different $D$-nucleus bound states are discussed in Sect.~\ref{sec3},
where we also compare with other microscopical models. The conclusions
are drawn in Sect.~\ref{sec4}.

 \section{The $D$-nucleus optical potential}
\label{sec2}
 
 In Ref.~\cite{Tolos:2009nn} the selfenergy of the $D$
 meson is evaluated in nuclear matter at various densities, $\rho$, as a
 function of the $D$ energy, $q^0$, and its momentum, $\vec q$, in
 the nuclear matter frame.  It is calculated by means of
\begin{eqnarray}
\Pi_D(q^0,\vec{q}\,;\rho)&= &\int_{p\le p_F} \frac{d^3p}{(2\pi)^3} \,
\left [~{T^\rho}^{(I=0,J=1/2)}_{DN}(P^0,\vec{P})+3 {T^\rho}^{(I=1,
J=1/2)}_{DN}(P^0,\vec{P})\right ] \ , \label{eq:pid}
\end{eqnarray}
 where $\vec p$ and $p_F$ are the momentum of the nucleon 
 and the Fermi momentum at nuclear density $\rho$, respectively. The quantity 
 $T^\rho_{DN}(P^0, \vec P )$ is 
 the  in-medium $DN$ $s$-wave interaction,
 with  total four-momentum  $(P^0,\vec P)$ in the 
 nuclear matter frame, 
 namely $P^0=q^0+E_N(\vec p)$ and $\vec P=\vec q + \vec p$.
 Here,   isospin symmetry is assumed, and
 the amplitude is summed over nucleons up to the Fermi level. 
 Since we are interested in finding bound states, 
 we shall be concerned about the $s$-wave $D$ selfenergy around the
 $DN$ threshold.
 
 The in-medium interaction $T^\rho_{D N}$ is obtained
 self-consistently including all coupled channels with the same
 quantum numbers.  In the work of Ref.~\cite{Tolos:2009nn}, the
 Bethe-Salpeter equation is solved with sixteen coupled channels for
 $I=0,~J=1/2$, twenty-two channels for $I=1,~J=1/2$, eleven channels
 for $I=0,~J=3/2$ and twenty for $I=1,~J=3/2$.  The medium effects in
 the scattering amplitude are the Pauli blocking on the intermediate
 nucleon states and the selfenergy of the charmed mesons ($D$ and
 $D^*$) in the intermediate channels. The pion and baryon selfenergies
 are not considered. While the baryon dressing did not change the
 qualitative behaviour of the $D$ meson in nuclear
 matter~\cite{tolos-angels-mizutani}, the coupling to intermediate
 states with pions is of minor importance for the $DN$ and $D^*N$
 dynamics in a dense environment \cite{Tolos:2009nn}.  Note the
 importance of the selfconsistency for $\Pi_D(q^0,\vec{q}\,;\rho)$,
 since the in-medium amplitude $T^\rho_{D N}$ contains the $D(D^*)N$
 channel, which depends in turn on $\Pi_{D(D^*)}(q^0,\vec{q}\,;\rho)$.

 \begin{figure}[t]
 \centerline{ 
 \epsfysize = 100 mm  \epsfbox{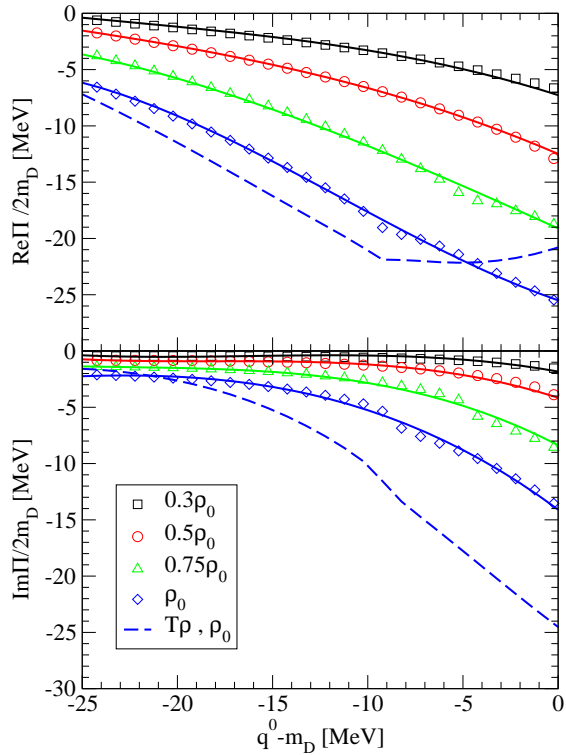}
              }
 \caption{ Selfconsistent $D$ selfenergy of
  Ref.~\cite{Tolos:2009nn} for zero momentum as a function of energy,
  and for four different densities. The solid lines stand for the fitted
  functions of Eqs.~(\ref{eq:self})-(\ref{eq:coef}). The dashed line
  shows the selfenergy for $\rho=\rho_0$ calculated in  
  the low density limit (without selfconsistency).
}
 \label{fig:iiabc}
\end{figure}

The $D$ selfenergy, scaled by $2 m_D$, is displayed with points in
Fig.~\ref{fig:iiabc} as a function of the $D$-meson energy around
threshold. It is shown for various nuclear medium densities, $\rho$,
and with the $D$ meson momentum fixed to zero. 

The $D$ selfenergy is evaluated in infinite nuclear matter. In finite
 nuclei we use the local density approximation (LDA), substituting
 $\rho$ by $\rho (r)$, which is the local density at each point in the
 nucleus taken from experiment. For the $s$-wave, as it is here the
 case, it was shown in Ref.~\cite{Nieves:ev} that LDA gave the same
 results as a direct finite nucleus calculation. Then,  the LDA
 $D$ selfenergy, $\Pi_D(q^0,r) \equiv \Pi_D(q^0,\vec 0,\rho(r))$
 allows to define an energy-independent local optical potential,
\begin{equation}
  V_{D}(r) = \frac{
  \Pi_{D}(q^0=m_D,\vec{q}=0,\rho(r))}{2 m_{D}},
\label{eq:UindepE}
\end{equation}
from its  threshold ($q^0=m_D,~\vec{q}=0$) value.  This
 prescription gives a potential of $(-25 -i14)$~MeV at normal nuclear
 matter density $\rho_0=0.17\ {\rm fm}^{-3}$ as can be read off from
 Fig.~\ref{fig:iiabc}.  This would mean that one can expect bound states with
 approximately $-20$~MeV binding and a half-width of about 14~MeV.

 However, both the real and the imaginary parts of the $D$ selfenergy,
 around the $D$-meson mass, show a pronounced energy dependence, as
 can be appreciated in Fig.~\ref{fig:iiabc}. For instance,
 the real part at $q^0-m_D=-20$~MeV is about one forth of its value at
 $q^0=m_D$.  Hence, a realistic
 determination of the $D$ bound states should take this energy
 dependence into account.  Thus, we use an energy dependent optical
 potential defined as:
\begin{equation}
  V_{D}(r,E) = \frac{
  \Pi_{D}(q^0=m_D+E,\vec{q}=0,~\rho(r))}{2 m_{D}},
\label{eq:UdepE}
\end{equation}
where $E=q^0-m_D$ is the $D^0$ energy excluding its mass.  In order to
use the potential defined above in Eq.~(\ref{eq:UdepE}), the results of
Ref.~\cite{Tolos:2009nn} are parameterized in terms of analytical
functions in the energy range $-25 ~\mbox{MeV} < E < 0$ as (see
the solid lines in Fig.~\ref{fig:iiabc}),  
\begin{eqnarray}  
 \mbox{Re}[ \Pi_{D}(q^0=m_D+E, \vec 0~; \rho) ] &=&   
       a(\rho) + b(\rho)E 
     + c(\rho)E^2+ d(\rho)E^3 ,
\nonumber
\\
 \mbox{Im}[ \Pi_{D}(q^0=m_D+E, \vec 0~; \rho) ] &=&   
       e(\rho) + f(\rho)E 
     + g(\rho)E^2+ h(\rho)E^3 ,
\label{eq:self}
\end{eqnarray}
 with 
\begin{eqnarray}
 a(\rho)&=& (-84033.1   ~\rho/\rho_0    -25727.8 ~\rho^2/\rho_0^2 
+ 14536.2 ~\rho^3/\rho_0^3) ~\mbox{MeV}^2 ,
\nonumber
 \\
 b(\rho)&=& ( -8654.6  ~\rho/\rho_0 + 6475.6    ~\rho^2/\rho_0^2) ~\mbox{MeV} ,
\nonumber
 \\
 c(\rho)&=& -341.2   ~\rho/\rho_0 + 447.8  ~\rho^2/\rho_0^2 ,
\nonumber
 \\
 d(\rho)&=& (-5.648 ~\rho/\rho_0 + 8.777 ~\rho^2/\rho_0^2) ~\mbox{MeV}^{-1} ,
\nonumber
 \\
 e(\rho)&=& (-10616.5   ~\rho/\rho_0 -39532.6   ~\rho^2/\rho_0^2 
 ~-2489.6\rho^3/\rho_0^3) ~\mbox{MeV}^{2} ,
\nonumber
 \\
 f(\rho)&=& (-3356.6   ~\rho/\rho_0 -1337.44   ~\rho^2/\rho_0^2) ~\mbox{MeV} ,
\nonumber
 \\
 g(\rho)&=& -291.09   ~\rho/\rho_0 +  136.656  ~\rho^2/\rho_0^2 ,
\nonumber
 \\
 h(\rho)&=& ( -6.70 ~\rho/\rho_0 + 5.20~\rho^2/\rho_0^2)
 ~\mbox{MeV}^{-1} . 
\label{eq:coef}
\end{eqnarray}
In the next section we will solve the Schr\"odinger equation, with
both the energy dependent [Eq.~(\ref{eq:UdepE})] and 
independent [Eq.~(\ref{eq:UindepE})] potentials, to find bound
states for different nuclei through the periodic table. Since the 
$D$-meson optical potential is much smaller than its mass, we expect
the relativistic corrections to be tiny and certainly smaller than the
theoretical uncertainties of the interaction. We will also
discuss the implications of our results in the practical search for
these $D^0$ bound states.

 \section{Results}
\label{sec3}

We look for $D^0$-nucleus bound states  by  solving the
 Schr\"odinger equation:
 \begin{equation}
   \left[ -{{\vec\nabla^2}\over{2 \mu}} +  V_{\rm{opt}}(r) \right] \Psi \,
  = (B-i \Gamma /2) \Psi ,
\label{eq:SchE}
 \end{equation}
where $B$ is the binding energy ($B<0$), $\Gamma/2$ the half-width of
the bound state and $\mu$ is the $D$-nucleus reduced mass.  As
mentioned above, we will present results from two different potentials
$V_{\rm{opt}}(r)=V_D(r,E=B)$ and $V_{\rm{opt}}(r)=V_D(r,E=0)$.
 
 We solve the Schr\"odinger equation in coordinate space by using a
 numerical algorithm~\cite{Oset:1985tb,GarciaRecio:1989xa}, which has
 been extensively tested in similar problems of
 pionic~\cite{Nieves:ev, Nieves:1991ye} and antikaonic~\cite{baca}
 atomic states and in the search of possible antikaon~\cite{baca} and
 $\eta$~\cite{GarciaRecio:2002cu} nucleus bound states.  Charge
 densities are taken from Ref.~\cite{Ja74}.  For each nucleus, we take
 the neutron matter density approximately equal to the charge one,
 though we consider small changes, inspired by Hartree-Fock
 calculations with the DME (density-matrix expansion)~\cite{Ne75} and
 corroborated by pionic atom analysis~\cite{nog92}.  In Table~1 of
 Ref.~\cite{baca} all the densities used throughout this work can be
 found.  However, charge (neutron matter) densities do not correspond
 to proton (neutron) ones because of the finite size of the proton
 (neutron).  We take that into account following the lines of
 Ref.~\cite{Nieves:ev} and use the proton (neutron) densities in our
 approach.

 \begin{figure}[t]
\begin{center}
 \includegraphics[width=0.5\textwidth,angle=-90]{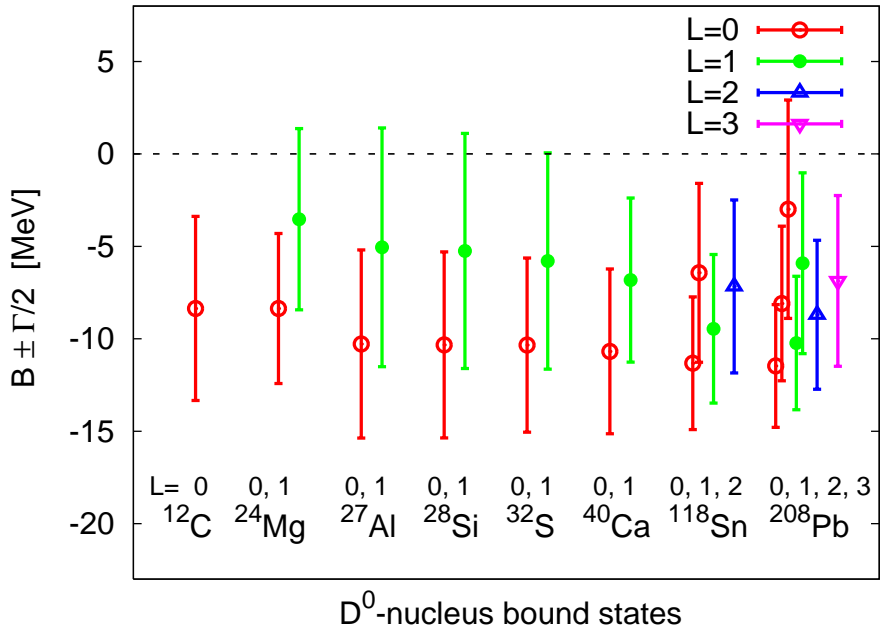}
\caption {\small Binding energies and widths for different $D^0$-nucleus 
  states obtained using the strong energy dependent $D-$potential.}
\label{fig:BdepE}
\end{center}
\end{figure}

 Results, binding energies and widths, from the energy dependent
 potential are shown in Table~\ref{table:BdepE}.  These results are
 also presented in Fig.~\ref{fig:BdepE}, where the states
 found for different nuclei and orbital angular momentum are
 collected.  The bound states for which the absolute value of the binding
 energy is much smaller than the corresponding half width have not been
 presented, because they mix with the continuum energy spectrum and do
 not define clear bound states.  For instance, in $^{40}$Ca we also
 find a $2s$ state with a binding energy of $-1.3$~MeV and half-width of
 6.7~MeV, which mixes with the continuum. We have disregarded
 the consideration of this state and others alike.

 From the results of Table~\ref{table:BdepE}, we conclude there exist
 chances to see distinct peaks corresponding to $D^0$ bound states.
 Let us consider angular momentum $l=0$. For medium size nuclei, up to
 $^{40}$Ca, there exists only one relevant $D^0$ bound state in each
 nucleus, with a binding energy around $-10$~MeV and half-width
 roughly below 5~MeV, which can be subject to experimental
 detection. For heavier nuclei, like $^{118}$Sn and $^{208}$Pb, we
 find two $l=0$ bound states, with an energy separation similar to
 their half-widths (around $4$~MeV).  In the case of angular momentum
 $l=1$, and for medium size nuclei (from $^{24}$Mg up to $^{32}$S)
 we find only the  $1p$ state, with $|B|$ smaller or equal than
 the $\Gamma/2$. Thus, these states are in the edge of the possible
 experimental determination. For heavier nuclei like $^{40}$Ca and
 $^{118}$Sn there is a well defined $1p$ bound state. In the case of
 $^{208}$Pb, we find the $1p$ and $2p$ bound states separated 
 by 4.3 MeV, while their half-widths are 3.6 and 4.9
 MeV, respectively . We also find states with $l=2$ for $^{118}$Sn and
 $^{208}$Pb, and $l=3$ for $^{208}$Pb.  

\begin{figure}[t]
\begin{center}
\makebox[0pt]{\includegraphics[width=0.4\textwidth,height=0.3\textheight]{func.eps}\hspace{1.5cm}
\includegraphics[width=0.4\textwidth,height=0.25\textheight]{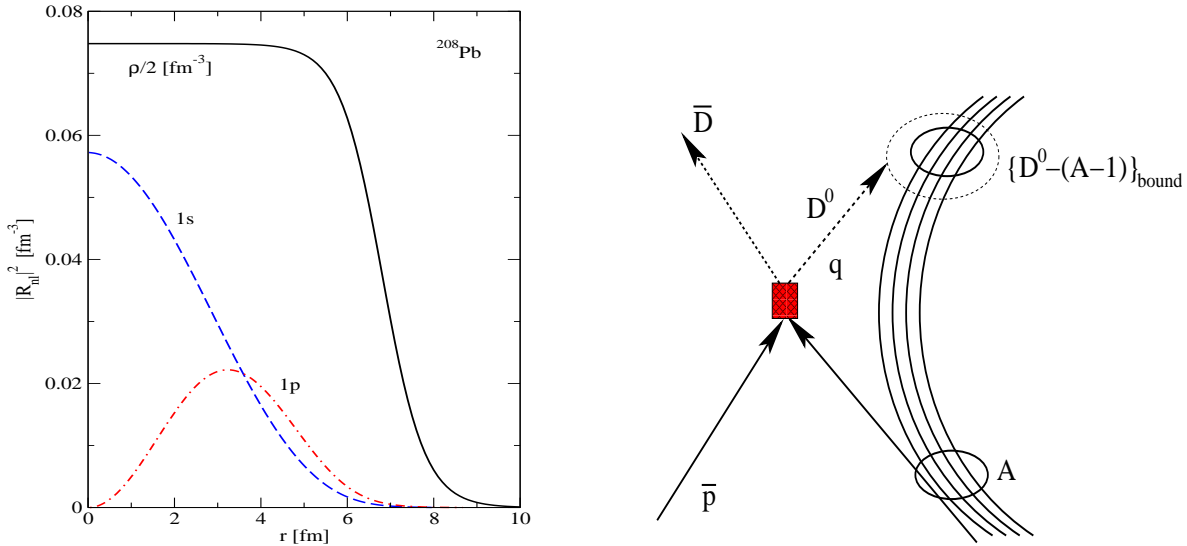}}
\caption {\small Left: Squared absolute value of the radial part of the wave
  function of the $1s$ and $1p$ levels in $^{208}$Pb. The nuclear
  density for lead is also shown. Right: Possible mechanism for production of
  $D^0$ mesic nuclei using a beam of antiprotons.}
\label{fig:foPb.trap}
\end{center}
\end{figure}

In summary, for light and medium nuclei (from $A=12$ up to $A=40$ ),
there are observable $D^0$ mesic nuclei $1s$ states. For heavier
nuclei, and in addition to the $1s$ level, we also find clearly
observable $1p$ states assuming that different angular momentum can be
separated.  For all cases, the bound $D^0$ meson is not orbiting
around the nucleus, but rather it is embedded inside of it. For
instance this can be seen in the left panel of
Fig.~\ref{fig:foPb.trap}, where the squared absolute values of the
radial wave function for the $1s$ and $1p$ levels in $^{208}$Pb,
together with the nuclear density, are shown.

To clarify the most relevant aspects of the model, we have also
obtained the bound state spectrum when some of the ingredients of the
full model are not considered. In Table~\ref{table:BindepE}, we show
results for $^{12}$C, $^{40}$Ca and $^{208}$Pb nuclei with different
approximations. First, we examine the effects produced by the strong
energy dependence of the optical potential and have computed energies
and widths with the energy independent optical potential of
Eq.~(\ref{eq:UindepE}). We see that when the energy dependence of the
potential is neglected, the states become more bound by a factor
ranging from 1.6 in $^{12}$C to 2 in $^{208}$Pb, and also the widths
of the states became larger by approximately a factor of two. These
results can be easily understood by looking at the energy dependence
of the optical potential in Fig.~\ref{fig:iiabc}. There, we see that
both, the real (attractive) and imaginary (negative) parts of the
optical potential decrease, in absolute value, by a factor larger than 2
from the threshold, $q^0=m_D$, to $q^0=m_D-15$~MeV. Thus, the energy
dependence of the optical potential plays a major role and need to be
taken into account.

Next, we examine in Table~\ref{table:BindepE} the importance of the
selfconsistency in the calculation. To this end, we have recalculated
the full spectrum by using an energy dependent optical potential
deduced from the $D$ selfenergy in the low density limit,
$\Pi_{\rm{low}}\sim T\rho$, where $T$ is the $ND$ T-matrix in free
space. This amounts to use $T^{\rho =0}$, instead of $T^\rho$ in
Eq~(\ref{eq:pid}). In Fig.~\ref{fig:iiabc}, we also compare the $D$
selfenergy calculated for normal nuclear matter density with and
without selfconsistency for zero momentum and energies below
threshold.  Though, the real part of the $\Pi_D(q^0,\vec{0},\rho)$
selfconsistent selfenergy does not differ much from that of the low
density theorem approach, however the absolute values of the imaginary
part of the $\Pi_{\rm{low}}$ approach are around a 60\% larger than
the exact result of $\Pi_D(q^0,\vec{q}=0,\rho)$ for $\rho=\rho_0$, and  in
the energy region $-20~$MeV~$\leq q^0-m_D\leq 0$. As a consequence,
the obtained bound states (see Table~\ref{table:BindepE}) using the
low density limit have larger widths and smaller binding energies (due
to the repulsive effect of the larger imaginary part) than those
calculated with the exact selfconsistent potential.

After analyzing the relevance of the ingredients of the model, we also
compare our results in $^{208}$Pb with those obtained in
Ref.~\cite{tsushima}. There,  a relativistic mean field calculation is
carried out that leads to binding energies  much larger than those obtained
here. For instance, the $1s$ state in $^{208}$Pb is bound by almost one
hundred MeV and has no width in the model of Ref.~\cite{tsushima},
whereas we find a binding of only  about ten MeV with a width of 6.6~MeV
for the same $1s$ level (see Table~\ref{table:BindepE}).  Of course,
our coupled channel unitary and selfconsistent model is much more
elaborated and it is also able to predict decay widths.

\begin{table}[t]
\caption{ ($-B$, $\Gamma/2$) in MeV for $D^0-$nucleus bound states 
calculated with the energy dependent selfenergy.}
\footnotesize
\vspace*{.4cm}
{
\begin{tabular}{c|cccccccc}
\hline
\hline \jtstrut
  &   $^{12}$C  & $^{24}$Mg    &  $^{27}$Al    &  $^{28}$Si & $^{32}$S   &   $^{40}$Ca  & $^{118}$Sn &  $^{208}$Pb    \\
\hline \jtstrut
1s&(7.0, 5.0)&(8.4, 4.1)&(10.3, 5.1)&(10.3, 5.0)&(10.3, 4.7)&(10.7, 4.5)&(11.3, 3.6)&(11.5, 3.3)\\
1p&            &(3.5, 4.9)&( 5.0, 6.5)&( 5.3, 6.4)&( 5.8, 5.9)&( 6.8, 5.4)&( 9.5, 4.0)&(10.2, 3.6)\\
1d&             &            &               &               &     &              &( 7.2, 4.7)&( 8.7, 4.0)\\
2s&             &              &               &               &     &            &( 6.4, 4.8)&( 8.1, 4.2)\\
1f&             &              &               &               &     &              &        &( 6.9, 4.6)\\
2p&             &              &               &               &     &              &        &( 5.9, 4.9)\\
\hline
\hline
\label{table:BdepE}
\end{tabular}
}\normalsize
\end{table}
\begin{table}[t]
\begin{center}
\caption{ ($-B$, $\Gamma/2$) in MeV for $D^0-$nucleus bound states in
  $^{12}$C, $^{40}$Ca and $^{208}$Pb. The first set of results have
  been obtained by using the energy independent potential of
  Eq.~(\ref{eq:UindepE}). In the second set of results, though the
  energy dependence of the optical potential is kept, binding energies
  and widths have been obtained from a low density optical potential,
  where effects due to the selfconsistency treatment have been
  neglected. The last column gives, for comparison, the predicted
  binding energies of Ref.~\cite{tsushima} for
  $^{208}$Pb (in MeV).} 
\vspace*{.4cm} 
\label{tbl:statedep2}
\footnotesize
\begin{tabular}{c|ccc|ccc|c}
\hline
\hline \jtstrut
 &\multicolumn{3}{|c|}{Energy independent $V_{\rm opt}$}& 
 \multicolumn{3}{|c|}{Energy dependent $V_{\rm opt}$ without selfconsistency}& 
$-B$ of Ref.~\cite{tsushima}\\
\hline \jtstrut
  &   $^{12}$C   &   $^{40}$Ca   &  $^{208}$Pb & $^{12}$C  &   $^{40}$Ca   & $^{208}$Pb & $^{208}$Pb\\
\hline \jtstrut
1s&(11.1, 10.9)&(18.4, 13.5)&(20.2, 11.0)& (7.5, 10.2)&(12.2, 7.5)&(13.0, 5.2)& 96.2\\
1p&            &(11.0, 11.4)&(17.9, 10.7) &           &(7.3, 11.1) &(6.6, 9.5)& 93.0\\
1d&              &               &(15.0,10.3)& & &(2.6, 12.8)& \\
2s&              &               &(13.9,10.1)& & & (1.5, 13.7)& 88.5 \\
1f&              &               &(11.7,9.8 )& & & & \\
2p&              &               &(9.9,9.5) & & & \\
\hline
\hline
\label{table:BindepE}
\end{tabular}
\normalsize
\end{center}
\end{table}

Finally, we  address the possibility of finding $D^+$ mesic nuclei.
The calculated $D-$nucleus strong optical potential is the same for 
$D^0$ and $D^+$ mesons because of the isospin symmetry. 
The found full optical potential is not very deep. The strength of the
 attractive potential is, at most, 20~MeV (see Fig.~\ref{fig:iiabc}).  
 A positive charged $D^+$ meson also
 feels the nuclear attraction of the nuclei. However, the Coulomb
 repulsion due to the positive electric charge of the nuclei is
 important, especially for heavy nuclei.  We have considered a light
 nuclei, $^{12}$C, and studied the $D^+$-nucleus bound states by
 adding the Coulomb repulsive potential to the optical one. We have
 found that the $1s$ state has a binding energy of $B=-4.6$~MeV and a
 half-width of $\Gamma/2=6.4$~MeV which is larger that $|B|$. Hence
 this state is not a clear case for experiments because it has a
 significant overlap with the energies of the continuous energy
 spectrum. Obviously for heavier nuclei, with more protons, there is
 not a better chance of finding $D^+-$nucleus bound states.  We do not
 study the case of nuclei lighter than $^{12}$C, because the LDA used
 here is not reliable enough for small nuclei, specially when
 significant cancellations among Coulomb and optical potential are
 taking place.  Summarizing, with the optical potential considered in
 this work, there are no chances of finding $D^+$ mesic nuclei with
 atomic number $Z$ equal or larger than 6.

In the coupled channel calculation of Ref.~\cite{Tolos:2009nn}, used as
input for the $D$ meson optical potential of this work, 
the nuclear matter vector meson $D^*$ selfenergy was
calculated as well. The in-medium $D^*$ selfenergy is found there to be
repulsive for energies around the threshold and densities $\rho\leq
\rho_0$, hence that model predicts that there are not $D^*$-nucleus
bound states.

\begin{figure}[t]
\begin{center}
\makebox[0pt]{\includegraphics[width=0.4\textwidth,height=0.35\textheight]
{selfD_rho.eps}\hspace{1.5cm}
\includegraphics[width=0.3\textwidth,height=0.25\textheight]{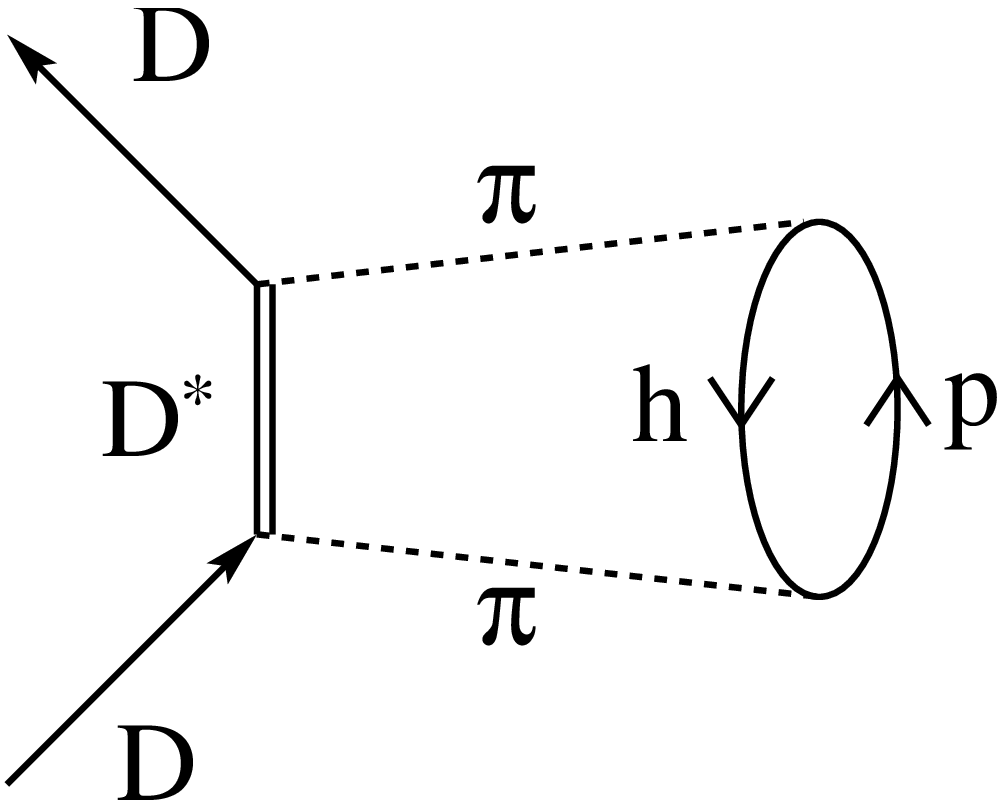}}
 \caption{ Left: Comparison of the $\Pi_D$ selfenergy over $2m_D$
  versus nuclear density for different values of the meson energy
  $q^0= m_D$ and $q^0= m_D-$15~MeV for SU(8) and SU(4) models. Right:
  Contribution to the $D-$selfenergy induced by the $p-$wave $D^* \to
  D \pi$ and $NN\pi$ vertices.} \label{fig:su4}
\end{center}
\end{figure}

The SU(8) model used here as a kernel in the coupled channel unitary
calculation reduces to the SU(4) model for $J=1/2$, when the vector
meson coupling constants are set to zero.  To establish the relevance
of the inclusion of the vector mesons, specially of the $D^*$, in the
$D$ meson dynamics, we have also performed the calculations with the
SU(4) model. In the left panel of Fig.~\ref{fig:su4}, both the SU(8)
and SU(4) $D$ selfenergies are displayed for comparison. There, it can
be appreciated that at threshold ($q^0=m_D$) and for $\rho\leq\rho_0$,
the SU(4) potential is small and repulsive, while the SU(8) model
provides a small attraction. However, for $q^0-m_D=-15~$MeV and
$\rho\leq\rho_0$, both potentials are attractive. However, the
imaginary parts of both selfenergies are quite different, being much
larger for the SU(4) case (about $-60$~MeV for $\rho=\rho_0$) than for
the SU(8) model. This is due to the behaviour at finite density of the
different resonance-hole structures of the $D$-meson selfenergy close
to the $DN$ threshold in the SU(4)
\cite{tolos-angels-mizutani,tetsuro-angels} and SU(8)
\cite{Tolos:2009nn} models. The large imaginary part of the SU(4)
optical potential induces an effective repulsion, which together with
the small attraction of its real part leads to the no existence of
$D^0-$nucleus bound states within the SU(4) model. This is in contrast
to the SU(8) model that turns out to be attractive enough to give the
bound spectra shown in Table~\ref{table:BdepE} and
Fig.~\ref{fig:BdepE}.

A word of caution must be said here. The above discussion overlooks
the fact there will be some extra contributions to the $D-$selfenergy
from the $DN \to D^* N$ interaction mediated by pion exchange from the
$p-$wave $D^* \to D \pi$ and $NN\pi$ vertices (see right panel of
Fig.~\ref{fig:su4}). At threshold, these extra terms will be purely
real, since there is no phase space for the reaction $DN \to D^*N$,
and they would also affect  to the $s-$wave free space amplitudes derived in
Ref.~\cite{GarciaRecio:2008dp}. Because of the $p-$wave nature of the
involved vertices, we expect the $s-$wave part of 
these contributions to be small near threshold, and their
contribution to be effectively accounted for by the renormalization
parameters used in ~\cite{GarciaRecio:2008dp}, that were adjusted to
reproduced the lowest-lying $s-$wave charmed baryon resonances. Note,
however that these new mechanisms will be source of extra imaginary
part in the case of the $D^*-$selfenergy.
\section{Conclusions}
\label{sec4}

 We have used a recent $D^0-$nucleus optical potential, evaluated
 within a self-consistent unitarized coupled-channel approach to find
 $D^0-$nucleus bound states.  We have shown that  selfconsistency
 effects and the energy dependence of the optical potential play  major
 roles, and need certainly to be taken into account.  The potential is
 attractive and we find bound states for all studied nuclei through
 the whole periodic table.  On the other hand, it  produces states
 with  relatively small half widths, smaller in general than the binding
 energies, and in many cases smaller than the separation among levels,
 which makes possible their observation.

The strong energy dependence of the potential is due to the large
effect on the $D$ selfenergy of the charmed resonances, and their
medium modifications~\cite{Tolos:2009nn}, appearing close to $DN$
threshold~\cite{GarciaRecio:2008dp}. Taking into account the energy
dependence reduces the strength of both the real and imaginary parts
of the potential, below the $D^0$ threshold. This leads to
substantially narrower states, but at the same time also with smaller
binding energies.  The heavy nuclei accommodate many $D^0$-bound
states and the separation of the levels, for a fixed angular
momentum, is about half width of the states.  The best chances for
observation of bound states are in the region of $^{24}\mbox{Mg}$,
provided an orbital angular momentum separation can be done, where
there is only one $s-$ bound state and its half width is about a
factor of two smaller than the binding energy.  In any case we would
like to stress out that even if no broad bumps are found in the
experiments, they should find some strength in the bound region
stretching in energy down to the sum of the binding energy plus the
half width of the bound states. Short of having the values for the
binding energy and width of the states, this more limited information
is still very valuable to gain some knowledge on the $D$ nucleus
optical potential, and it should stimulate experiments in this
direction. This can be of particular interest for the future ${\rm
{\bar P}}$ANDA@FAIR physics program.  Nevertheless, we should point
out that the production and the experimental detection of these
$D^0-$nucleus bound states will likely be a quite difficult task.
One might think in reactions of the type (see right panel of
Fig.~\ref{fig:foPb.trap})
\begin{eqnarray}
 \bar p + A_Z &\to& D^- + \left\{ (A-1)_Z- D^0\right\}_{\rm bound}
 \nonumber \\
  &\to& \bar D^0 + \left\{ (A-1)_{Z-1}\,- D^0\right\}_{\rm
 bound}\label{eq:fair} 
\end{eqnarray} 
to be investigated in the future facility FAIR.  Since the above
reactions are two-body$\to$two-body, the outgoing ${\bar D}$ energy is
fixed, for a certain scattering angle in the laboratory system, and
the creation of the $D^0$ bound state will be identified as a peak in
the $d\sigma/d\Omega dE({\bar D})$ over a background of inclusive
(${\bar p}, {\bar D}$) cross section.  The High-Energy Storage Ring
(HESR) at FAIR running with full luminosity (limited by the production
rate of $2\times 10^7\, {\bar p}/s$) at momenta larger than 6.4~GeV
would produce about 100/s $D$ meson pairs around
$\psi(4040)$~\cite{PANDA}. Assuming that one per ten
million\footnote{This fraction should be understood only as an
educated guess. In reactions of the type of that of
Fig.~\ref{fig:foPb.trap}, the momentum transferred $q$ to the bound
$D^0$ is fixed. The cross section for the reaction is then
proportional to $|\Phi_{nlm}(q)|^2$, where $\Phi_{nlm}(q)$ is the
$D^0$ bound wave function in momentum
space~\cite{Nieves:1990xy,Nieves:1991yf}. Since in the reactions of
Eq.~(\ref{eq:fair}) the momentum transfer is quite large, around 2 GeV
at threshold, and at least 1 GeV for incoming antiproton momenta of 10
GeV or larger, one can immediately see that the cross sections will be
small, since the $D^0$ bound wave function has difficulty to
accommodate such large momenta; but the reaction is possible. Given
the typical size of the $D^0$ bound states (see for instance the left
panel of Fig.~\ref{fig:foPb.trap}), momentum transfers of about 0.2
GeV can be easily accommodated, and in these circumstances one might
expect that one per cent of the $D^0$ mesons could be
trapped~\cite{Nieves:1990xy,Nieves:1991yf}. The extra six orders of
magnitude of reduction, implicit in the fraction $10^{-7}$, would
account for the expected large reduction of the $D^0$ bound wave
function when the momentum increases from 0.2 GeV to 1 GeV (we have
assumed a factor $10^{-3}$).} of the produced $D^0$ mesons is trapped
in a bound orbit, we are then left with a production rhythm of around
only one event every $10^5$ seconds or equivalently to few hundred
events per year.

Recoilless production reactions have
proved to be more efficient in the case of detecting deeply pionic
bound states~\cite{Nieves:1991yf,Hirenzaki:1991us,Itahashi:1999qb}. One
possible reaction, where negligible momentum transfers could be
achieved, is
\begin{equation}
D^{*+} + A_Z \to \pi^+ + \left\{ A_Z- D^0\right\}_{\rm bound}.
\end{equation}
From the theoretical point of view we would expect sizeable formation
peaks over a flat background.  However, since the $D^*$ is an unstable
particle, it will not be possible to create a $D^*$ beam, which makes,
in practice, unfeasible the above reaction\footnote{ Around ten
millions of $\bar{D}^{(*)}D^{(*)}$ pairs per year are expected to be
produced at BESIII in the region of the $\psi(3770)$ and $X(4160)$
resonances~\cite{Zou}. Even if it were possible to put a thin nuclear
target of 1 cm$^2$ at a distance at small as 1 meter from the
collision area, we would not have more than one per year charmed meson
interacting with the nuclear target.}.  The above discussion brings us
to reconsider the reactions of Eq.~(\ref{eq:fair}), and creating in
the primary $\bar p N$ collision, not only a $D\bar{D}$ pair, but also
a virtual pion, which will induce many body modes (particle-hole) in
the nuclear medium~\cite{Hernandez:1992rv}. Such modes will carry
almost no energy, but high momentum components, which would allow the
virtual $D^0$ meson in the right panel of Fig.~\ref{fig:foPb.trap} to
have a significantly smaller momentum, being in this way,
significantly increased the $D^0$ bound state production cross
section. The main drawback would be that one is not facing now with a
two-body going to two-body reaction. Hence to guaranty the creation of
the $D^0-$nucleus bound state, it would be needed to look at the decay
products, $\Lambda_c \pi$ and $\Sigma_c \pi$ pairs, after the
absorption of the $D^0$ by the nucleus.  However, once the decay
products of $D^0-$nucleus bound state need to be detected, one
realizes that it is unnecessary to use antiprotons (secondary beam) as
projectiles, and instead one can benefit from the use of protons
(primary beam), obtaining in this way a large enhancement factor
($\sim 10^5$) in the incoming flux. Thus, one could look at reactions
of the type
\begin{equation}
p  + A_Z \to p + {\bar D}^0 + \underbrace{\left\{ A_Z- D^0\right\}_{\rm
    bound}}_{\begin{array}{cc}\hookrightarrow &\Lambda_c \pi + X \\
&\Sigma_c \pi + X  \end{array}}
\end{equation}
Given the fact that there are more that two particles in the final state,
it is possible to pick up regions of the phase space where the
momentum, that needs to be accommodated in the $D^0$ bound wave
function, would be sufficiently small to make significantly larger the 
probability of $D^0$ trapping. Moreover, the above reaction is
coherent; there is no change of charge in $p+N \to p + {\bar D}^0
+D^0+ N$ and the final nuclear state is the same as the initial
one. This provides a factor $A^2$ in the cross section versus a factor
$A$ in the inclusive reaction which gives the
background~\cite{Nieves:1991yf}. Because of that, $D^0-$bound states
in heavier nuclei than magnesium, mentioned above, might have
better chances to be detected.

 Finally, we would like to point out that in this work we have used a
 model for the in medium $D-$meson selfenergy that, it is based on a
 SU(8) spin-flavor extension~\cite{GarciaRecio:2008dp} of the SU(2)
 Weinberg-Tomozawa (WT) pion-nucleon $s-$wave interaction in the free
 space\footnote{Of course, both flavor and spin symmetry breaking
 terms are included in the model of Ref.~\cite{GarciaRecio:2008dp} to
 account for the physical hadron masses and meson decay
 constants.}. When vector mesons are not considered, our in medium
 selfenergy reduces, up to minor details, to that deduced in
 Ref.~\cite{tetsuro-angels}. This latter one  is based on a SU(4) flavor
 extension of the SU(2) vacuum WT pion-nucleon $s-$wave
 interaction. The optical potential deduced from
 Ref.~\cite{tetsuro-angels} is not attractive enough to bind $D^0$
 nuclear states. This is in sharp contrast with our findings here. One of
 the major differences between both approaches,  and mostly responsible
 for this distinctive difference, is that the model used here treats
 heavy pseudoscalar and heavy vector mesons on equal footing, as
 required by Heavy Quark Symmetry (HQS).  This latter symmetry is a proper
QCD  spin-flavor symmetry~\cite{IW89,Ne94,MW00} when the
quark masses become much larger than the typical confinement scale,
$\Lambda_{\rm QCD}$.

\section*{Acknowledgments}
  We warmly thank B.S. Zou, A. Gillitzer and E. Oset for useful
 discussions. This work is partly supported by DGI and FEDER funds,
 under contract FIS2008-01143/FIS, the Spanish Ingenio-Consolider 2010
 Program CPAN (CSD2007-00042), the Junta de Andalucia grant
 no. FQM225, and Generalitat Valenciana under contract
 PROMETEO/2009/0090. We acknowledge the support of the European
 Community-Research Infrastructure Integrating Activity "Study of
 Strongly Interacting Matter" (acronym HadronPhysics2, Grant Agreement
 n. 227431) under the Seventh Framework Programme of EU.  Work
 supported in part by DFG (SFB/TR 16, "Subnuclear Structure of
 Matter"). L.T. acknowledges support from the RFF program of the
 University of Groningen and the Helmholtz International Center for
 FAIR within the framework of the LOEWE program by the State of Hesse
 (Germany).

\end{document}